\documentclass[12pt]{article}
\begin{document}
\def\bfone{\relax{\rm 1\kern-.35em 1}}\let\shat=\hat
\def\hat{\widehat}

%
\def\cS{{\cal K}}
\def\IE{\relax{{\rm I\kern-.18em E}}}
\def\cE{{\cal E}}
\def\rt{{\cR^{(3)}}}
\def\IGam{\relax{{\rm I}\kern-.18em \Gamma}}
\def\IGa{\IA}
\def\cV{{\cal V}}
\def\Rt{{\cal R}^{(3)}}
\def\tft#1{\langle\langle\,#1\,\rangle\rangle}
\def\IA{\relax{\hbox{{\rm A}\kern-.82em {\rm A}}}}
\def\hata{{\shat\a}}
\def\hatb{{\shat\b}}
\def\hatA{{\shat A}}
\def\hatB{{\shat B}}
\def\bv{{\bf V}}
\def\Fb{\overline{F}}
\def\nablab{\overline{\nabla}}
\def\Ub{\overline{U}}
\def\Db{\overline{D}}
\def\zb{\overline{z}}
\def\eb{\overline{e}}
\def\fb{\overline{f}}
\def\tb{\overline{t}}
\def\Xb{\overline{X}}
\def\Vb{\overline{V}}
\def\Cb{\overline{C}}
\def\Sb{\overline{S}}
\def\delb{\overline{\del}}
\def\Gammab{\overline{\Gamma}}
\def\Ab{\overline{A}}
\def\Anh{A^{\rm nh}}
\def\alphab{\bar{\alpha}}
\def\cy{Calabi--Yau}
\def\cabg{C_{\alpha\beta\gamma}}
\def\B{\Sigma}
\def\Bh{\hat \Sigma}
\def\Kh{\hat{K}}
\def\Knh{{\cal K}}
\def\A{\Lambda}
\def\Ah{\hat \Lambda}
\def\R{\hat{R}}
\def\V{{V}}
\def\T{T}
\def\Gammah{\hat{\Gamma}}
\def\twot{$(2,2)$}
\def\K{K\"ahler}
\def\rat{({\theta_2 \over \theta_1})}
\def\lv{{\bf \omega}}
\def\w{w}
\def\CP{C\!P}
\def\o#1#2{{{#1}\over{#2}}}
\def\eq#1{(\ref{#1})}
\thispagestyle{empty}
\begin{titlepage}
\thispagestyle{empty}
\begin{flushright}
CERN-TH/98-48\\
UCLA/98/TEP/4\\
February 1998\\
\end{flushright}
\vskip 2.cm
\begin{center}
{\Large\bf Horizon Geometry, Duality and Fixed scalars in Six
 Dimensions\renewcommand{\thefootnote}{\fnsymbol{footnote}}\footnote{Work
 supported in part by EEC under TMR contract ERBFMRX-CT96-0045 (LNF Frascati,
 Politecnico di Torino and Univ. Genova), Angelo Della Riccia fellowship,
 CONICIT fellowship and by DOE grant DE-FG03-91ER40662}}
\vskip 2.cm
{\large L. Andrianopoli$^{a,c}$, R. D'Auria$^ b$, S. Ferrara$^c$,
 M.A. Lled\'o$^{d,e}$.}
\end{center}
{\it $^a$ Dipartimento di Fisica, Universit\`a di Genova, via Dodecaneso 33,
 I-16146 Genova and Istituto Nazionale di Fisica Nucleare (INFN)-Sezione di
 Torino, Italy.}

{\it $^b$ Dipartimento di Fisica, Politecnico di Torino, Corso Duca degli
 Abruzzi, 24, I-10129 Torino and Istituto Nazionale di Fisica Nucleare
 (INFN)-Sezione di Torino, Italy.}

{\it $^c$ CERN Theoretical Division, CH 1211 Geneva 23, Switzerland.}

{\it $^d$  Centro de F\'{\i}sica. Instituto Venezolano de Investigaciones
 Cient\'{\i}ficas (IVIC). Apdo 21827 Caracas 1020-A. Venezuela.}

{\it $^e$ Physics Department, University of California, Los Angeles. 405
 Hilgard Av. Los Angeles, CA 90095-1547. USA.} 
\begin{abstract}
We consider the problem of extremizing the tension for BPS strings in $D=6$
 supergravities with different number of supersymmetries. General formulae for
 fixed scalars and a discussion of degenerate directions is given. Quantized
 moduli, according to recent analysis, are supposed to be related to conformal
 field theories which are the boundary of three dimensional anti-de Sitter
 space time.
\end{abstract}
\end{titlepage}
\section{Introduction}
Recently possible connections of brane dynamics of the world volume theory and
 supergravity in the ambient space of
 near horizon geometry have been proposed and investigated
 \cite{jm, ckp, ss, hyun, sb, bps, mal}. These attempts heavily rely on
 the correspondence of supergravity on an anti-de Sitter bulk  and special
 properties of ``topological" singleton
 representations confined on anti-de Sitter boundaries \cite{gm}.

Interestingly enough,  as pointed out by Maldacena \cite{jm}, this
 correspondence can only occur at some special points of the moduli space of
 the theory. Indeed this observation is based on previous recent analysis
 which shows that certain moduli fields satisfy, in a given black brane
 background, flow equations \cite{fks} with an attractor point at the horizon
 where the geometry becomes
 anti-de Sitter \cite{gt}.

Fixed scalars (attractors) have been until recently mainly investigated in
 connection to black holes in $D=4$ and black holes and black strings in
 $ D=5$.

In $ D=5$ strings are dual (magnetic) to black holes so fixed scalars in these
 cases are easily obtained for both strings and black holes by using dual
 formulae replacing the BPS mass of a black hole with the BPS tension of a
 string \cite{ckr}. However, the recent developments on possible dualities 
 between anti-de Sitter spacetimes and brane dynamics has led to focus on
 string horizon geometries at $D=6$ because of their relation to two
 dimensional conformal field theories \cite{jm, ss, bps, ms}.

In the present paper we generalize the analysis of fixed scalars to $D=6$ BPS
 strings, a case which has not been thoroughly treated in previous
 investigations. We will examine cases with 8, 16 and 32 supersymmetries and
 explicitely compute fixed scalars for all these cases. A discussion of
 particular theories to which these results apply will also be given.

\section{Central charges and BPS tension in $D=6$ supergravity theories}

We recall here some properties of the moduli dependence of BPS string tension 
in six dimensional theories.

At $D=6$, 3-form field strengths can be self-dual or antiself-dual. Let us denote
 by $P$, $Q$ the number of self-dual and antiself-dual field strengths
 respectively.

For all $D=6$ theories we can use a unified formalism based on an underlying
  moduli space relevant to the problem, locally of the form
 $\hbox{O}(P,Q)/\hbox{O}(P)\times\hbox{O}(Q)$, where the range of $P,Q$ will
 depend on the theory \cite{r,s1}. More specifically for theories with (1,0)
 supersymmetry $P=1$, $Q=n$. For (1,1) supersymmetry $P=Q=1$, for (2,0)
 supersymmetry $P=5$, $Q=n$ and for (2,2) theories $P=Q=5$. Moreover, since
 (1,0) and (2,0) theories are chiral, $n$ is constrained by the anomaly
 cancellation to be $n=21$ in (2,0) theories and $n=(273+n_V-n_H)/29$ in (1,0)
 theory ($n_H, n_V$ being the number of hyper and vector multiplets
 respectively)\footnote{ The construction of D=6 chiral theories has been
 recently completed in  \cite{frs}}.

The coset space $\hbox{O}(P,Q)/\hbox{O}(P)\times \hbox{O}(Q)$ can be defined
 by coset representatives
 $X_{r\Lambda}, X_{I \Lambda}, r=1,\dots P, I=1,\dots Q, \Lambda =1,\dots P+Q)$
 satisfying the conditions 
\begin{equation}
X_{r\Lambda}X_{r\Sigma}-X_{I\Lambda}X_{I\Sigma}=\eta_{\Lambda\Sigma}.
\end{equation}

Explicitly, $X_{rI}$ are $PQ$ coordinates and: 
\begin{equation}
X_{rs}=\sqrt{\delta_{rs}+X_{rI}X_{sI}};\qquad X_{IJ}=\sqrt{\delta_{IJ}+
X_{rI}X_{rJ}}.
\end{equation}

To define BPS string tension through the BPS condition, we must introduce a 
``metric" $\mathcal{N}_{\Lambda \Sigma}$ for the self-dual and antiself-dual
 3-forms $H^{\Lambda}
$. The metric $\mathcal{N}_{\Lambda \Sigma}$ is defined as follows \cite{adf1}
\begin{equation}
\mathcal{N}_{\Lambda\Sigma}=X_{r\Lambda}X_{r\Sigma}+X_{I\Lambda}X_{I\Sigma}=
2X_{r\Lambda}X_{r\Sigma}-\eta_{\Lambda \Sigma}.
\end{equation}
with the property:
\begin{equation}
  \label{propn}
  {\cal N} \eta {\cal N}\eta  = \bfone \, , \quad \mbox{i.e.}\quad {\cal N}^{-1} = \eta {\cal N}\eta
\end{equation}
The self-duality condition of the field strengths becomes:
\begin{equation}
\mathcal{N}_{\Lambda\Sigma}{^*}\!H^{\Sigma}=\eta_{\Lambda\Sigma}H^{\Sigma}
\label{dual}
\end{equation}
which therefore implies, by integrating Eq. \ref{dual} on a 3-sphere
\begin{equation}
q_\Lambda=\eta_{\Lambda\Sigma}m^\Sigma
\end{equation}
$q_\Lambda, m^\Lambda$ being the electric and magnetic charges respectively.

Therefore, $P$ forms have $q=m$ and $Q$ have $q=-m$ as appropriate for self-dual
 (antiself-dual) forms. Note that the Dirac-Zwanzinger-Schwinger
 quantization condition for dyonic odd p-forms 
is \cite{adf1}\cite{deser}
\begin{equation}
q_\Lambda m'^\Lambda+m^\Lambda q'_\Lambda=2\pi k
\end{equation}
Using the equation above we get the manifestly $\hbox{O}(P,Q)$ invariant
 condition
\begin{equation}
q_\Lambda\eta^{\Lambda\Sigma}q'_{\Sigma} =\pi k
\end{equation}

We note that chiral theories, with $P\neq Q$, are non-lagrangian because of
a net number of self- (antiself-) dual tensors \cite{r},\cite{s1}.
However for non-chiral theories, with $P=Q$, Eq. \ref{dual} can be replaced
by $P$ unconstrained 2-form fields for which an action exists.
In an appendix we give the relation between the lagrangian couplings of
 non-chiral theories and the matrix ${\cal N}_{\Lambda\Sigma} $ appearing in
Eq. \ref{dual}.

From the quantized set of charges $q$ we must construct ``dressed" charges,
 appropriate to discuss BPS states. They are defined as follows,
\begin{equation}
Z_r=X_r^{\Lambda} q_\Lambda,\qquad Z_I=X_I^{\Lambda} q_\Lambda
\end{equation}
with the property 
\begin{equation}
W=Z_rZ_r+Z_IZ_I=q^{\Lambda}\mathcal{N}_{\Lambda\Sigma} q^\Sigma,\qquad
 Z_rZ_r-Z_IZ_I=q^{\Lambda}\eta_{\Lambda\Sigma} q^\Sigma\label{weinhold}
\end{equation}

Note that the first of the above equations defines an analogous of the
 Weinhold potential \cite{fgk} for $D=6$, BPS strings.
The physical meaning of $Z_r$ and $Z_I$ is that they provide the 3-form charge
 eigenstates which enter in the self-dual tensor
of the gravitational sector and the antiselfdual matter tensor of (1,0) and 
(2,0) theories respectively.
 For the (1,1) and (2,2) theories $Z_r$ and $Z_I$ all belong to the gravity
 multiplet.
 Let us note that $Z_r$ (and not $Z_I$) is a ``central charge" which occurs in 
the chiral supersymmetry algebra,
 while in the non chiral case both $Z_r$ and $Z_I$ are central charges.

For (1,0)  theory we have one $Z$ \footnote{In the (1,1) theory
 $Z^2=Z_L^2=Z_R^2 +em$ where $Z_L$ ($Z_R$) is the central charge of the (1,0)
 ((0,1)) subalgebra and $e,m$ denote the electric and magnetic charges of the
 BPS string. Both 1/2 and 1/4 BPS strings exist in this case.} while for (2,0)
 theories $Z_r$ is a 5 of $\hbox{USp}(4)$. This means that in both cases there
 is only one central charge eigenvalue and only one type of BPS state 
(1/2 BPS).

For the (2,2) theory, $Z_r$ and $Z_I$ are in the 5's of two different
 $\hbox{USp}(4)$. Therefore, there are two independent central charge
 eigenvalues and 1/4 BPS states are those for which $Z_r^2 \neq Z_I^2$
 \footnote{ By $Z_r^2$ we mean $Z_rZ_r$}
 while 1/2 BPS correspond to light-like charges 
$q^\Lambda\eta_{\Lambda\Sigma}q^{\Sigma}=0$ \cite{fm}.

We also remark that the (1,0) geometry of tensor multiplets is entirely
 analogous to the D=5 ``very special" geometry \cite{gst} for vector multiplets
 in respect to the fact that to each tensor multiplet corresponds one real 
scalar as for vectors in D=5.
It is not then surprising that the central charge and the moduli geometry
 carries a resemblance from that case.

First observe that the potential is now
\begin{equation}
W=Z^2+Z_I^2 \,  , \quad \, Z = X^\Lambda q_\Lambda
\end{equation}
where 
\begin{equation}
Z_I=P^i_I\partial_iZ
\end{equation}
and $P^i_I$ is the inverse Vielbein of $\hbox{O}(1,n)/\hbox{O}(n)$. Then it
 follows that 
\begin{equation}
Z_I^2=G^{ij}\partial_iZ\partial_jZ
\end{equation}
and in particular the condition $Z_I=0$ is equivalent to $\partial_iZ=0$.

From the previous equations we also note the relation
\begin{equation}
G_{ij}=\partial_iX^\Lambda \mathcal{N}_{\Lambda\Sigma}\partial_jX^\Sigma,
\qquad (X^\Lambda\eta_{\Lambda\Sigma}X^\Sigma=1)
\end{equation}
which follows from the Maurer--Cartan equations:
\begin{equation}
\partial_i X_\Lambda = X_{\Lambda I} P^I_{,i}
\end{equation}
and the relation:
\begin{equation}
{\cal N}^{\Lambda \Sigma} X_{\Sigma I} = X^\Lambda_I
\end{equation}

The explicit expressions of $G_{ij}$ and $ \mathcal{N}_{\Lambda\Sigma}$ are
\begin{eqnarray}
G_{ij}&=&\delta_{ij}-{x_ix_j\over 1+x_i^2},\qquad x_0=\sqrt{1+x_i^2},\qquad
 x^i=x_i
\\
{\cal N}_{\Lambda\Sigma}&=&2X_\Lambda X_\Sigma-\eta_{\Lambda\Sigma}
\label{enne}
\end{eqnarray}
where Eq. \ref{enne} follows from:
\begin{eqnarray}
X_\Lambda X_\Sigma - X_{\Lambda I} X_{\Sigma I} &=& \eta_{\Lambda \Sigma} 
\nonumber\\
X_\Lambda X_\Sigma + X_{\Lambda I} X_{\Sigma I} &=& {\cal N}_{\Lambda \Sigma}
\end{eqnarray}

\section{Fixed scalars and Horizon geometry}

From general considerations based on the relation of brane dynamics to
 asymptotic horizon geometry we expect \cite{adf1} that the flow of moduli
 toward the horizon implies an ``attractor" condition equivalent to the
 extremization of $W$ as defined by Eq. \ref{weinhold}.

This condition signals ``supersymmetry enhancement" in the horizon geometry
 \cite{fk, fgk}. It is remarkable that such ``phase transition" occurs at 
``special points" of the ``asymptotic" moduli space of the theory.

In D=4,5 black holes and strings the asymptotic geometry is
 $\hbox{Ad}_2\times\hbox{S}_2$ and  $\hbox{Ad}_3\times\hbox{S}_2$ respectively
 \cite{gt}.

Supersymmetry enhancement of the former, the so called Bertotti--Robinson
 geometry, is due to the fact that the background is superconformal invariant
 in a 4-dimensional sense \cite{kp}. For D=5 the geometry of the string 
corresponds to a (4,0) two dimensional conformal field theory with the
  supersymmetry enhancement in the corresponding superalgebra.

In these theories the ``fixed moduli" are the vector multiplets while the
 hypermultiplets are not fixed by the horizon geometry \cite{fks, fk}.

In the D=6 case, according to the analysis  of Ref. \cite{adf1}, fixed scalars
 correspond to partners of antiselfdual tensors, i.e. the moduli of (1,0)
 tensor multiplets. In this case the horizon geometry is
 $\hbox{Ad}_3\times\hbox{S}_3$.

For theories with higher supersymmetry (i.e. with P=5) the degenerate
 directions should correspond to D=6  (1,0) hypermultiplets, in analogy with
 the phenomenon in D=4,5 for theories with $N>2$ supersymmetry \cite{adf2}.
 Fixed scalars can be obtained following the analysis of Ref.\cite{adf2} for
 D=4,5 theories.

The basic point are the Maurer--Cartan equations, satisfied by the one form
 differentials of the BPS central charges \cite{adf1}.

For a $\hbox{O}(P,Q)/\hbox{O}(P)\times\hbox{O}(Q)$ moduli geometry these
 equations read

\begin{equation}
\nabla Z_r=P_r^IZ_I\label{mc1}
\end{equation}
\begin{equation}
\nabla Z_I=P_I^rZ_r
\end{equation}
where $P_{rI}$ is the Vielbein one-form ($P_r^I=\delta^{IJ}P_{rI},
 P^r_I=\delta^{rs}P_{sI}$). From the first equation we get that the
 ``central charge" $Z_r$ is extremized by $Z_I=0$, which therefore solves
 $\delta W=0$. It is important to notice that ``fixed scalars" for which
 $Z_I=0$
 give a ``regular" horizon since in this  case the analogous of the  
Bekenstein--Hawking entropy
\begin{equation}
S={A^{D-2}\over G_N} 
\end{equation}
takes the value
\begin{equation}
W|_{Z_I=0}=q^\Lambda\eta_{\Lambda \Sigma}q^{\Sigma}
\end{equation}
which therefore requires the charge vector $q$ to satisfy
 $q^\Lambda \eta_{\Lambda\Sigma}q^\Sigma =q^2 >0$ when 
$P<Q$.
For $P=Q$, the solution  with $q^2 < 0$  is isomorphic to the previous one by
 interchanging $Z_r $ with $Z_I$.

We now consider the details of the general remarks to the four particular 
theories with 8,16,32 supersymmetries \footnote{The (1,1) theory, having one 
tensor field, is like the (1,0) theory for $n=1$. However the string is 1/4
 BPS when it is dyonic \cite{dfk}, while it is 1/2 BPS when it is purely
 electric or magnetic.}.

In the (1,0) theory the ``attractor equation":
\begin{equation}
\partial_iZ=0
 \end{equation}
has the unique solution
\begin{equation}
X_\Lambda={q_\Lambda\over \sqrt{q^2}}
\end{equation}
The string tension at this point is
\begin{equation}
Z|_{extr}=\sqrt{q_\Lambda q^\Lambda}=\sqrt{q_0^2-q_i^2}
\end{equation}
Note that the hypermultiplet scalars (which must be present in the theory
 because of anomaly cancellation) do not enter in the discussion so the
 3-dimensional anti-de Sitter geometry cannot depend on them \cite{gm}.

For $q^2<0$, the equation $\delta W =0$ has a different solution which occurs 
for tensionless strings, i.e. at the point $Z=0$.
This is not an extremum of the string tension, but rather at this point:
\begin{equation}
  W\vert_{Z=0} = |Z_I|^2 = -q^2
\end{equation}
A particular solution of this equation is:
\begin{equation}
  X^{(0)}_i = \frac{q_i}{\sqrt{q_i^2}}\frac{q_0}{\sqrt{q_i^2 -q_0^2}}
\end{equation}

Let us now consider the (2,0) theory with $P=5, Q=21$. From the form of $W$
 and Eq. \ref{mc1} we see that 
\begin{equation}
\partial_iW=0
\label{attractor}\end{equation}
occurs at $Z_I=0$ if $q^2 >0$.

Since in this case $i$ takes  $5Q$ values and $Z_I=0$ are $Q$ conditions, it
 means that there are precisely $4Q$
  (84 for the case at hand) moduli directions not fixed by Eq. \ref{attractor}.
 These are precisely the Q
 hypermultiplets in the (1,0) decomposition of the (2,0) tensor multiplets.

Note that, according to the results of Ref. \cite{adfft}, the relevant solvable
 algebra decomposition of the coset reads 
\begin{equation}
\hbox{solv}\bigl({\hbox{O}(5,Q)\over \hbox{O}(5)\times\hbox{O}(Q)}\bigr)=
\hbox{solv}\bigl({\hbox{O}(1,Q)\over \hbox{O}(Q)}\bigr)+\hbox{solv}
\bigl({\hbox{O}(4,Q)\over O(4)\times\hbox{O}(Q)}\bigr)
\end{equation}
where the two factors correspond to the (1,0) tensor- and hyper-multiplets
 respectively.

Eq. \ref{attractor} implies that the hypermultiplets do not occur in the
 solution so  $4Q$ directions are undetermined by the attractor condition and
 will not occur  at the extremum of $W$.

It is straightforward to give a particular solution for the fixed scalars
 which solves Eq. \ref{attractor}. This is given by
\begin{equation}
X^{(0)}_{r\Lambda}={q_rq_\Lambda\over \sqrt{q_r^2}\sqrt{q_r^2-q_I^2}}
\end{equation}
It can be easily shown that at $X_{rI}=X^{(0)}_{rI}$
\begin{equation}
Z_r={q_r\sqrt{q_r^2-q_I^2}\over \sqrt{q_r^2}}
\end{equation}
which implies
\begin{equation}
W|_{X=X^{(0)}}=Z_r^2=q_r^2-q_I^2
\label{fixed}\end{equation}
Note that $X^{(0)}$ is a particular solution of $Z_I=0$ since it does not
 determine the moduli directions $\hat X$ orthogonal to $Z_r$
\begin{equation}
\hat X_{rI}=X_{rI}-{Z_rX_{sI}Z_s\over Z_s^2}
\end{equation}
Using Eq. \ref{enne} it is also possible to give a general formula for the
 ${\cal N}_{\Lambda \Sigma}$ metric at the attractor point. It reads
\begin{equation}
\mathcal{N}_{\Lambda \Sigma}|_{X=X^{(0)}}={2q_\Lambda q_\Sigma\over q^2}-
\eta_{\Lambda\Sigma}\end{equation}
which is of course consistent with Eq. \ref{fixed}.

Let us now consider the case $q^2 <0$.
Similar to the (1,0) case, $\delta W=0$ occurs in this case at $Z_r=0$
at which point:
\begin{equation}
  W|_{Z_r=0} =Z_I^2 =-q^2
\end{equation}
This equation implies that tensionless strings are not extrema of the BPS tension.
A particular solution of the moduli at this point is:
\begin{equation}
  X^{(0)}_{rI}=\frac{q_I}{\sqrt{q_I^2}}\frac{q_r}{\sqrt{q^2_I-q^2_r}}
\end{equation}
This solution leaves $5(Q-1)=100$ directions undetermined as it appears from 
the moduli directions orthogonal to $Z_I$: 
\begin{equation}
  \hat X_{rI} =X_{rI}- \frac{Z_I X_{rJ}Z_J}{Z_J^2}
\end{equation}

We finally come to the (2,2) theory with R-symmetry given by
 $\hbox{O}(5)\times\hbox{O}(5)\sim \hbox{USp}(4)_L\times\hbox{USp}(4)_R$.
 In this case the moduli space is locally $\hbox{O}(5,5)/
\hbox{O}(5)\times\hbox{O}(5)$ and the attractor condition
\begin{equation}
Z_I=0
\end{equation}
implies that the lower eigenvalue of the central charge matrix vanishes, as in
 the $D=4,5$ cases \cite{fk,adf2}. For charges  
$q^\Lambda q_\Lambda \, {^{<}_{>}}\,\,  0$, $Z_r^2 \,{^{<}_{>}}\,\, Z_I^2$,
 and the ``attractor point" gives a 
1/4 BPS state with Weinhold potential given by
\begin{eqnarray}
W|_{Z_I=0}=q^2 \quad &\mbox{if}&\quad q^2>0 \\
W|_{Z_r=0}=-q^2 \quad &\mbox{if}&\quad q^2<0
\end{eqnarray}
This situation corresponds to the BPS orbit $\hbox{O}(5,5)/\hbox{O}(4,5)$
 according to the analysis of Ref. \cite{fm, lps}.

One half BPS strings require, on the other hand, two coinciding eigenvalues for
 the central charge and this requires a light-like orbit
 $q^\Lambda q_\Lambda=0$. The BPS orbit is in this case $\hbox{O}(5,5)/
\hbox{IO}(4,4)$ and implies (for all values of the moduli) the identity
\begin{equation}
Z_r^2=Z_I^2
\end{equation}

Note that in the previous cases of chiral theories the 1/2 BPS orbits fall
 into three categories, i.e. $\hbox{O}(P,Q)/\hbox{O}(P-1)\times\hbox{O}(Q)$ for
 $q^2>0$, $\hbox{O}(P,Q)/\hbox{IO}(P-1,Q-1)$ for $q^2=0$ and
 $\hbox{O}(P,Q)/\hbox{O}(P)\times\hbox{O}(Q-1)$ 
for $q^2<0$. Light-like orbits
 do not correspond to supersymmetry enhancement in this case. This is similar 
 to 1/8 BPS orbits with vanishing entropy for D=4 black holes \cite{fm, fg}.
Space--like orbits ($q^2<0$) correspond to the occurrence of tensionless strings \cite{w}.
The moduli at which the string becomes tensionless do not correspond to
an extremum of the string tension.

\section{Concluding remarks}

In this paper we have considered in some details BPS strings in different
 theories at $D=6$ and studied the extremization of the string tension in the
 moduli space.

Fixed scalars correspond to particular quantized values in terms of the string
 charges.

In (1,0) theories the fixed scalars are in tensor multiplets, in (1,1) theory
 the fixed scalar is  in the gravity multiplet (dilaton) and in (2,0), (2,2)
 theories the fixed scalars correspond to the  tensor multiplet scalars
 resulting  by decomposing the theories in (1,0) representations. The other
 scalar fields are not fixed by the horizon geometry, therefore 84 and 20
 moduli are at their initial value in the (2,0) and the (2,2) theories
 respectively.

The four kinds of $6D$ supergravity theories considered here  correspond,
 in string theory language, to different compactifications of heterotic and 
Type II strings on $\hbox{K}_3$ and $\hbox{T}_4$. (1,0) theories with more
 than one tensor multiplet naturally arise in open string constructions
 \cite{s2} and F--theory compactifications \cite{v}. The fixed scalars
 correspond to special isolated points in the $\hbox{K}_3$ and $\hbox{T}_4$
 moduli spaces. These points should play an important role because it is
 precisely at these values that the horizon geometry $\hbox{Ad}_3\times
 \hbox{S}_3$ is defined.

3-d anti-de Sitter supergravity has been recently reconsidered on the new
 light of possible duality relations between anti-de Sitter physics and the
 brane (string in this case) conformal dynamics \cite{ss,bps, ms}.

\section{Acknowledgements}

One of us (S.F.) would like to thank J. Maldacena and K. Sfetsos for useful 
discussions.

\appendix
\section{Appendix}
In this appendix we consider the relation of the ``coupling''
 ${\cal N}_{\Lambda\Sigma}$ in Eq. \ref{dual} and the lagrangian of $P$
 unconstrained 2-form potentials.

The basic lagrangian is:
\begin{equation}
  \label{lagr}
  \frac{1}{2} g_{\Lambda\Sigma} H^\Lambda \wedge {}^* H^\Sigma +
 \frac{1}{2} b_{\Lambda\Sigma} H^\Lambda \wedge  H^\Sigma 
\end{equation}
where $g_{\Lambda\Sigma}$ is the real symmetric kinetic coupling and
$b_{\Lambda\Sigma}$ is the real antisymmetric ``axionic'' coupling.

Decomposing :
\begin{equation}
  \label{dec}
  H^\Lambda =H^{+\Lambda}+ H^{-\Lambda}
\end{equation}
in self-dual and  antiself-dual parts we get:
\begin{equation}
  \label{selflag}
\frac{1}{2}{\cal N}_{+\Lambda\Sigma}H^{+\Lambda}\wedge H^{-\Sigma} +
\frac{1}{2}{\cal N}_{-\Lambda\Sigma}H^{-\Lambda}\wedge H^{+\Sigma}
\end{equation}
with:
\begin{equation}
  \label{defn}
  {\cal N}_{\pm\Lambda\Sigma}= \mp g_{\Lambda\Sigma}+
b_{\Lambda\Sigma}
\end{equation}
and therefore:
\begin{equation}
  {\cal N}_+ = -{\cal N}_-^T
\end{equation}
${\cal N}_\pm$ transforms with a fractional transformation under the $O(P,P)$
 action, i.e.:
 \begin{equation}
{\cal N}_\pm^\prime = \left( C+D {\cal N}_\pm \right)
 \left( A+B {\cal N}_\pm\right)^{-1}
\label{trasfn} 
 \end{equation}
where:
\begin{eqnarray}
  \left(\matrix{ A&B\cr C&D \cr }\right)&\in & O(P,P):\nonumber\\
A^TC +C^TA = B^TD+D^TB =0 &,& A^TD+C^TB =1
\end{eqnarray}
The relation between $ {\cal N}_\pm$ as given by Eq. \ref{defn} and the matrix
${\cal N}$ as given in Eq. \ref{dual} turns out to be:
\begin{eqnarray}
  \label{reln}
  {\cal M} \equiv C^T {\cal N} C & = & \left(\matrix{
2({\cal N}_-^{-1} -{\cal N}_+^{-1})^{-1} & 
({\cal N}_- +{\cal N}_+)({\cal N}_- -{\cal N}_+)^{-1} \cr
- ({\cal N}_- -{\cal N}_+)^{-1}({\cal N}_- +{\cal N}_+) &
2({\cal N}_- -{\cal N}_+)^{-1}\cr}\right)\nonumber\\
&=&\left(\matrix{g-bg^{-1}b & bg^{-1} \cr -g^{-1}b & g^{-1} \cr}\right)
= \left(\matrix{1&b\cr 0 &1\cr}\right) \left(\matrix{g&0\cr 0&g^{-1}\cr}\right)
\left(\matrix{1&0\cr -b &1\cr}\right)
\end{eqnarray}
where:
\begin{equation}
  C=\frac{1}{\sqrt{2}}\left(\matrix{1&1\cr - 1 & 1\cr}\right).
\end{equation}
The matrix given by Eq. \ref{reln} satisfies the property:
\begin{equation}
  {\cal M}\hat \eta {\cal M}\hat \eta = \bfone
\end{equation}
with respect to the off-diagonal metric:
\begin{equation}
  \hat \eta = \left(\matrix{0&1\cr 1&0\cr}\right)= C^T \eta C
\end{equation}

The above equation, Eq. \ref{reln}, is obtained using the results of ref. \cite{gazu}
for the parametrization of the coset representative $X$ of $O(P,P)/[O(P) \times
O(P)]$ as explained in ref. \cite{adf1}.

In other words, the $2P\times 2P$ matrix ${\cal N}$ transforms as a $O(P,P)$
tensor when ${\cal N}_\pm$ undergo the projective transformation given by
Eq. \ref{trasfn}.


\end{document}